\documentclass[preprint]{article}
\usepackage{graphicx}
\usepackage{dcolumn}
\usepackage{bm}
\usepackage{amsmath}
\usepackage{amssymb}
\usepackage{multirow}
\usepackage{caption}
\usepackage{epstopdf}
\usepackage{hyperref}
\usepackage{cite}

\begin{document}
{\setlength{\oddsidemargin}{1.2in}
\setlength{\evensidemargin}{1.2in} } \baselineskip 0.55cm
\begin{center}
{\LARGE {\bf Quintessence Star Solutions with Conformal Symmetry in a Durgapal Spacetime}}
\end{center}
\date{\today}
\begin{center}
  Meghanil Sinha*, S. Surendra Singh \\
Department of Mathematics, National Institute of Technology Manipur,\\
Imphal-795004,India\\
Email:{ meghanil1729@gmail.com, ssuren.mu@gmail.com}\\
 \end{center}
 
\textbf{Abstract}: The accelerated expansion of the Universe can be suitably attributed to the existence of the dark energy (DE). On the backdrop of this concept, this paper introduces a novel, anisotropic compact star model whose stability and structure are governed by the presence of quintessence field, defined by the parameter $ w_{Q} (-1<w_{Q}<-\frac{1}{3}) $ and which admits conformal symmetry. The construction of the model relied on the Durgapal-Fuloria (DP) metric formulation. The model successfully meets all the necessary physical constraints viz., TOV equation, energy conditions, compactness factor, surface redshift and casuality condition. The results are analyzed through analytical methods as well as through the graphical visualization for the various physical attributes.\\


\section{Introduction}\label{sec1}

\hspace{0.8cm}Undeniably, General Relativity (GR) stands as the most ground-breaking theoretical achievement in the last century \cite{R1,R2,R3}. The theory of Special Relativity first demonstrated Einstein's brilliance before the world, and it was the generalization of this theory that resulted in GR \cite{R4}. Using intricate tensor calculations, Einstein revealed the mysteries and beauty of the spacetime within the Universe \cite{R5}. Prior to this revolutionary work, tensor analysis was relegated to being just a mathematical tool. The masterful intellect of him also introduced the concept of gravitational waves (GW). Einstein field equations (EFE) are crucial for understanding and explaining a wide range of astrophysical phenomena. A century after G-R was introduced, we can now successfully detect GW, using advanced instruments like LIGO and VIRGO. Essentially, this provided astrophysicists and cosmologists a new arena for future research. Almost a century ago, Schwarzschild provided the very first mathematical description of EFEs for a compact, self-gravitating celestial body \cite{C1}. It provided the initial mathematical model for describing the gravitational field around any static, symmetric, spherical object. While the model has its limitations, it served as a crucial starting point, compelling the researchers to find more precise solutions to EFEs that govern massive objects. As a result, we now possess a large number of exact solutions to the field equations that describe a very vast array of stellar objects, while a few demonstrate physical compatibility, but there might be some bounds, while others don't \cite{C2,M3,K,DP,BG,DP00,DP01,DP02}. A large collection of the existing exact solutions were derived by postulating specific conditions for the spacetime geometry or the internal matter properties \cite{M1,M2,KP2,O,OD}. Spherical symmetry is the standard assumption for modelling various static stars.\\
The current astrophysical and cosmological era is defined by two major problems: the accelerated expansion driven by DE and the missing mass represented by dark matter (DM) \cite{S1,S2,S3}. According to the hypothesis, GR necessitates DE, a form of exotic energy to explain the Universe as we observe it today \cite{S4,S5}. The observed facts about the accelerating expansion and the existence of DM have created a major theoretical challenge to GR \cite{RT1,RT2,RT3}. The expansion of the Universe is a central and crucial phenomenon currently commanding the attention of modern cosmology and astrophysics. Past theories suggested that the Universe's expansion would cease one day, but current evidence reveals that it is expanding at an ever-increasing rate instead. Measurements from Cosmic Microwave Background (CMB) indicate that the Universe in overwhelmingly composed of DE $(68.3 \%)$, DM $(26.8 \%)$ and ordinary matter (OM) $(4.9 \%)$ \cite{RG1}. DM exerts an attractive gravitational force on matter; although it cannot be seen directly, its presence is inferred by observing its gravitational effects. The existence of DE can be deduced by measuring the large-scale wave patterns found in the Cosmos. DE is distinguished by having a strong negative pressure, implying its equation of state (EOS) is negative. The EOS of DE is defined as $ p = w \rho, ( \omega < -\frac{1}{3} ) $. It has been the subject of extensive investigation by numerous researchers \cite{VT1,VT3,VT4}. Choosing $ w = -1 $ yields the gravastar model, which is also categorized as DE star \cite{RL,RL9,RL10,MG,M4}. A value of $ w < -1 $ corresponds to the phantom energy. Numerous authors have employed the concept of phantom energy as the necessary matter content to theorize the existence of a wormhole \cite{VT5,VT6,VT7,VT8}. Building on earlier studies, we have specifically selected quintessence DE as the foundation for developing our new model. The quintessence field is represented by defining the parameter $ w_{Q} $ where $ w_{Q} (-1 < w_{Q} < -\frac{1}{3}) $. We have modeled the internal fluid as a mixture of OM and a DE like matter that is inherently repulsive. Although they do not interact, but we have focused on modeling the total, cumulative impact of both components. The model assumes pressure inside the fluid as anisotropic with two distinct components, i.e. radial pressure $ (p_{r}) $ and transversal pressure $ (p_{t}) $. Here $ p_{t} $ represents the pressure perpendicular to $ p_{r} $ and $ \triangle = p_{t} - p_{r} $ is the anisotropic factor.\\
In particular, finding a solution for compact astrophysical objects that avoids a singularity has been a major, long-standing challenge in astrophysical community. Compact objects are generally understood to be various types of high-density bodies created at the end stage of a star's life cycle. Therefore, studying these stars requires investigating their microscopic features and how dense matter behaves in extreme environments - a task considered one of the most crucial in modern astrophysics. Clearly, we can see that getting a reliable description of dense compact objects remains a complex challenge, even though numerical investigations, experiments and observations have been conducted over the last few decades. Data collected from compact stars may provide the key information needed to reduce the major uncertainties surrounding the EOS at extremely high nuclear densities \cite{CVT1,CVT2,CVT3}. Exact solutions for the EFEs that describe static, spherical cosmic objects are increasingly seen as vital benchmarks that generate great interest among the mathematicians and physicists alike. Thus obtaining exact solutions for the field equations are necessary for progress in this area of study. The basic construction requires choosing a geometric formula (a metric ansatz) to start with and/or defining and EOS that relates the pressure and density of the fluid in our work. In our current research, we will specifically utilize the well known DP metric function for defining the spacetime \cite{DP1}. The choice of this specific metric potential is physically sound and well-behaved \cite{DP,DP2,DP3}. It is known to be reliable for producing singularity-free solutions for compact stellar entities that are positive and regular at their center \cite{DP4,DP5,DP6,DP7,DP8,DP9}. Hence, this approach offers a highly practical and effective way to model our star, and it also ensures a smooth and seamless connection between the star's interior and exterior manifolds.\\
Thus we are dedicating this investigation to a compact star model that incorporates quintessence and admits the symmetry known as conformal motion. Using the DP metric as our foundation, we then proceeded to examine the physical characteristics of the resulting model \cite{DP1,DP3,DP5,DP7}. The layout of the current paper is presented below : after a brief introduction in section (\ref{sec1}), we move on to section (\ref{sec2}), which details the interior solution and the EFEs. Section (\ref{sec3}) is dedicated to present the conformal killing vector and the solution derived using the DP metric. Some physical properties are investigated in section (\ref{sec4}), while section (\ref{sec5}) addresses the exterior spacetime and the necessary matching conditions. Section (\ref{sec6}) contains the TOV equations and section (\ref{sec7}) outlines the required energy conditions. Section (\ref{sec8}) contains a discussion of some stellar characteristics followed by the stability analysis presented in section (\ref{sec9}). And the paper closes with section (\ref{sec10}) which contains our conclusions and final thoughts.\\

\section{Field equation formalism}\label{sec2}

\hspace{0.8cm}We start our study by using the static, spherically symmetric line element to characterize the four dimensional spacetime as\\
\begin{equation}\label{1}
ds^{2} = -e^{m(r)}dt^{2} + e^{n(r)}dr^{2} + r^{2}(d \theta^{2} + \sin^{2}\theta d \phi^{2}).
\end{equation}\\
In this context, $ m(r) $ and $ n(r) $ are metric potentials - mathematical functions that depend only on the radial co-ordinate $ r $. Let us proceed by assuming that the stellar model incorporates both a quintessence-like field and having anisotropic pressure. The Einstein equations are formulated in the following way:
\begin{equation}\label{2}
G_{\alpha \beta} = 8\pi G ( T_{\alpha \beta} + \Gamma _{\alpha \beta} )
\end{equation}\\
where $ G_{\alpha \beta} = $ Einstein tensor and $ G = $ gravitational constant. The energy momentum tensor for the quintessence field, denoted as $ \Gamma_{\alpha \beta} $, is characterized by the parameter $ w_{Q} $ where $ (-1 < w_{Q} < -\frac{1}{3}) $. Considering the line element's signature, the components are stated as \cite{VT43}\\
\begin{equation}\label{3}
\Gamma_{t}^{t} = \Gamma_{r}^{r} = - \rho_{Q}
\end{equation}\\
\begin{equation}\label{4}
\Gamma_{\theta}^{\theta} = \Gamma_{\phi}^{\phi} = \frac{1}{2}(3 w_{Q} + 1)\rho_{Q}
\end{equation}\\
and the energy-momentum tensor is represented by\\
\begin{equation}\label{5}
T_{\alpha \beta} = (\rho + p_{r})u^{\alpha}u_{\beta} + p_{t}g^{\alpha}_{\beta} + (p_{r} - p_{t})\xi^{\alpha}\xi_{\beta}
\end{equation}\\
where $ u^{\alpha}u_{\beta} = - \xi^{\alpha}\xi_{\beta} = 1 $ and $ u^{\alpha}\xi_{\beta} = 0 $. Here $ u $ is the fluid's 4-velocity, $ \xi $ is the orthogonal space-like vector, $ \rho $ is the energy density and $ p_{r} $ and $ p_{t} $ are the radial and transverse pressures respectively. With the spacetime defined in equation (\ref{1}) along with the energy-momentum tensor, the EFEs simplify to\\
\begin{equation}\label{6}
\frac{e^{-n}n'}{r} - \frac{e^{-n}}{r^{2}} + \frac{1}{r^{2}} = 8 \pi (\rho + \rho_{Q})
\end{equation}\\
\begin{equation}\label{7}
\frac{e^{-n}}{r^{2}} + \frac{e^{-n}m'}{r} - \frac{1}{r^{2}} = 8 \pi (p_{r} - \rho_{Q})
\end{equation}\\
\begin{equation}\label{8}
e^{-n}\big[ \frac{m'^{2}}{2} + m'' - \frac{m'n'}{2} + \frac{m'-n'}{r} \big] =  8 \pi (p_{t} + \frac{3w_{Q} + 1}{2} \rho_{Q})
\end{equation}\\
where prime implies the derivative w.r.t $ r $.\\

\section{Mathematical framework from conformal killing equation and using the DP metric}\label{sec3}

\hspace{0.8cm}We use conformal killing vectors to find the inheritance symmetry, which allows us to study the inherent relationship between geometry and matter within the EFEs as\\
\begin{equation}\label{9}
\verb"L"_{\eta} g_{ab} = \Theta g_{ab}. 
\end{equation}\\ 
The symbol $ \verb"L" $ is the Lie derivative operator and $ \Theta $ is defined as the conformal vector being used in the operation. The specific type of vector symmetry is dictated by $ \Theta $, killing vectors require $ \Theta = 0 $, a constant value of $ \Theta $ yields a homothetic vector and if $ \Theta $ varies with position and time, it results in a conformal vector. Consequently, the conformal killing vector is a key tool for exploring the subtleties of spacetime geometry. The conformal killing equation is therefore expressed as\\
\begin{equation}\label{10}
\verb"L"_{\eta} g_{ab} = \eta_{a;b} + \eta_{b;a} = \Theta g_{ab}.
\end{equation}\\
When the line element from equation (\ref{1}) is applied, the following set of equations are derived\\
\begin{equation}\label{11}
\eta^{1}m' = \Theta
\end{equation}
\begin{equation}\label{12}
\eta^{4} = U
\end{equation}
\begin{equation}\label{13}
\eta^{1} = \frac{\Theta r }{2}
\end{equation}
\begin{equation}\label{14}
\eta^{1}n' + 2\eta^{1}_{,1} = \Theta. 
\end{equation}
Here $ U $ denotes a constant. The above equations lead to the following results\\
\begin{equation}\label{15}
e^{m(r)} = A^{2}r^{2}
\end{equation}
\begin{equation}\label{16}
e^{n(r)} = \big(\frac{B}{\Theta}\big)^{2}
\end{equation}
\begin{equation}\label{17}
\eta^{a} = U\delta^{a}_{4} + \big(\frac{\Theta r }{2}\big)\delta^{a}_{1}.
\end{equation}
The symbols $ A $ and $ B $ represent the integration constants. When these solutions are plugged into equations (\ref{6}-\ref{8}), we obtain the following results\\
\begin{equation}\label{18}
\frac{1}{r^{2}} - \frac{\Theta^{2}}{r^{2}B^{2}} - \frac{2 \Theta \Theta'}{rB^{2}} = 8 \pi (\rho + \rho_{Q})
\end{equation}
\begin{equation}\label{19}
\frac{3\Theta^{2}}{r^{2}B^{2}} - \frac{1}{r^{2}} = 8 \pi (p_{r} - \rho_{Q})
\end{equation}
\begin{equation}\label{20}
\frac{\Theta^{2}}{B^{2}r^{2}} + \frac{2 \Theta \Theta'}{rB^{2}} = 8 \pi (p_{t} + \frac{3w_{Q} + 1}{2} \rho_{Q}).
\end{equation}
We adopt the DP ansatz - a known formula for the metric potential $ e^{n(r)} $ in order to solve the equations. The formula is given by\\
\begin{equation}\label{21}
e^{n(r)} = \frac{7 + 14Nr^{2} + 7N^{2}r^{4}}{7 - 10Nr^{2} - N^{2}r^{4}}
\end{equation}\\
where $ N $ represents a constant which carries the dimension of $ [L^{-2}] $. The metric $ e^{n(r)} $ is clearly regular and well-defined at the center $(r=0)$. Combining equations (\ref{16}) and (\ref{21}) results in\\
\begin{equation}\label{22}
\Theta^{2} = \frac{B^{2}(7 - 10Nr^{2} - N^{2}r^{4})}{7 + 14Nr^{2} + 7N^{2}r^{4}}.
\end{equation}\\
Utilizing the expression for $ \Theta^{2} $ from equation (\ref{22}), in equations (\ref{18}-\ref{20}), we can now derive the following as\\
\begin{equation}\label{23}
\frac{8N(9 + 2Nr^{2} + N^{2}r^{4})}{7(1 + Nr^{2})^{3}} = 8 \pi (\rho + \rho_{Q})
\end{equation}
\begin{equation}\label{24}
\frac{-2(7 + 22Nr^{2} + 5N^{2}r^{4})}{7(r + Nr^{3})^{2}} = 8 \pi (p_{r} - \rho_{Q}) 
\end{equation}
\begin{equation}\label{25}
\frac{7 - 51Nr^{2} + 5N^{2}r^{4} - N^{3}r^{6}}{7r^{2}(1 +  Nr^{2})^{3}} = 8 \pi (p_{t} + \frac{3w_{Q} + 1}{2} \rho_{Q}).
\end{equation}\\
One key observation is that from equations (\ref{23}-\ref{25}) we have a system of three equations with four unknowns $ (\rho, p_{r}, p_{t} $ and $ \rho_{Q}) $. To find the solution, let us assume an E-O-S, where $ p_{r} $ is proportional to $ \rho $ as\\
\begin{equation}\label{26}
p_{r} = \varpi \rho, \hspace{0.5cm} 0 < \varpi < 1
\end{equation}\\
with $ \varpi = $ EOS parameter. Using the EOS parameter to close the system, we can now solve equations (\ref{23}-\ref{25}) and can find these resulting expressions as\\
\begin{equation}\label{27}
\rho = \frac{7 + 21Nr^{2} - 19N^{2}r^{4} - N^{3}r^{6}}{28(1 + \varpi)\pi r^{2}(1 + Nr^{2})^{3}}
\end{equation}
\begin{equation}\label{28}
p_{r} = \frac{\varpi(7 + 21Nr^{2} - 19N^{2}r^{4} - N^{3}r^{6})}{28(1 + \varpi)\pi r^{2}(1 + Nr^{2})^{3}}
\end{equation}
\begin{equation}\label{29}
\rho_{Q} = \frac{-\frac{-8N(9 + 2Nr^{2} + N^{2}r^{4})}{7(1 + Nr^{2})^{3}} + \frac{2(7 + 21Nr^{2} - 19N^{2}r^{4} - N^{3}r^{6})}{7(1 + \varpi)r^{2}(1 + Nr^{2})^{3}}}{8\pi}
\end{equation}
\begin{equation}\label{30}
p_{t} = \frac{-\frac{-(7 - 5Nr^{2} + 5N^{2}r^{4} - N^{3}r^{6})}{7r^{2}(1 + Nr^{2})^{3}} - \frac{1}{2}(1 + 3w_{Q})\Big(-\frac{8N(9 + 2Nr^{2} + N^{2}r^{4})}{7(1 + Nr^{2})^{3}} + \frac{2(7 + 21Nr^{2} - 19N^{2}r^{4} - N^{3}r^{6})}{7(1 + \varpi)r^{2}(1 + Nr^{2})^{3}}\Big)}{8\pi}.
\end{equation}\\
In this context, we have\\
\begin{equation}\label{31}
\rho^{eff} = \frac{N(9 + 2Nr^{2} + N^{2}r^{4})}{7 \pi (1 + Nr^{2})^{3}}
\end{equation}
\begin{equation}\label{32}
p_{r}^{eff} = \frac{7 - 22Nr^{2} - 5N^{2}r^{4}}{28 \pi (r + Nr^{2})^{3}}
\end{equation}
\begin{equation}\label{33}
p_{t}^{eff} = \frac{7 - 51Nr^{2} + 5N^{2}r^{4} - N^{3}r^{6}}{56 \pi r^{2}(1 + Nr^{2})^{3}}.
\end{equation}\\
\begin{figure}[ht!]
\centering
\includegraphics[scale=0.5]{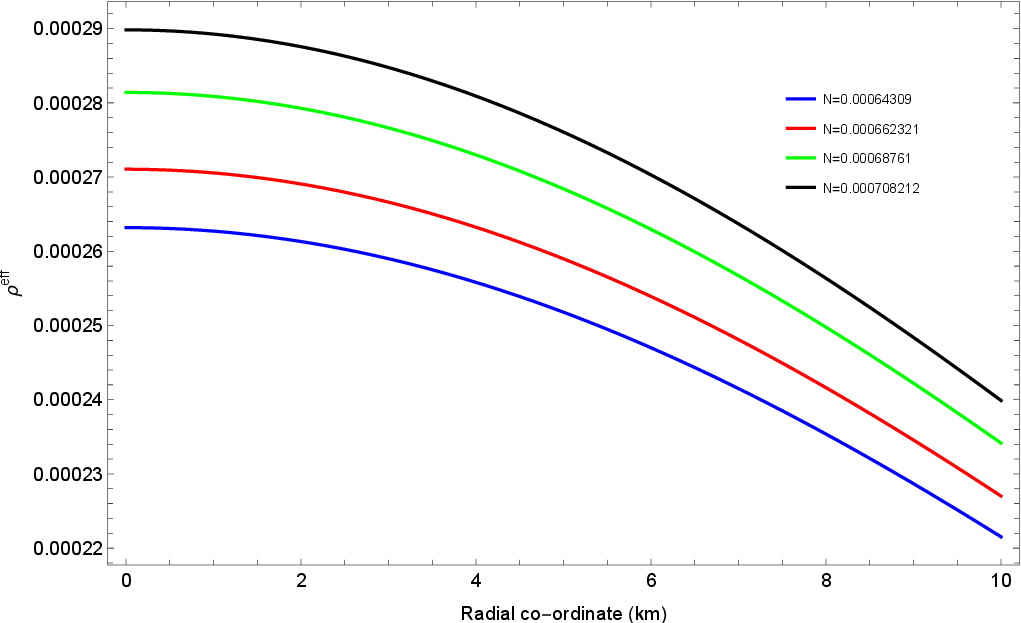}
\caption{Plot of $ \rho^{eff} $ versus radial co-ordinate for $ N = 0.00064309, N = 0.000662321, N= 0.00068761, N = 0.000708212 $ }\label{1}
\end{figure}\\
\begin{figure}[ht!]
\centering
\includegraphics[scale=0.5]{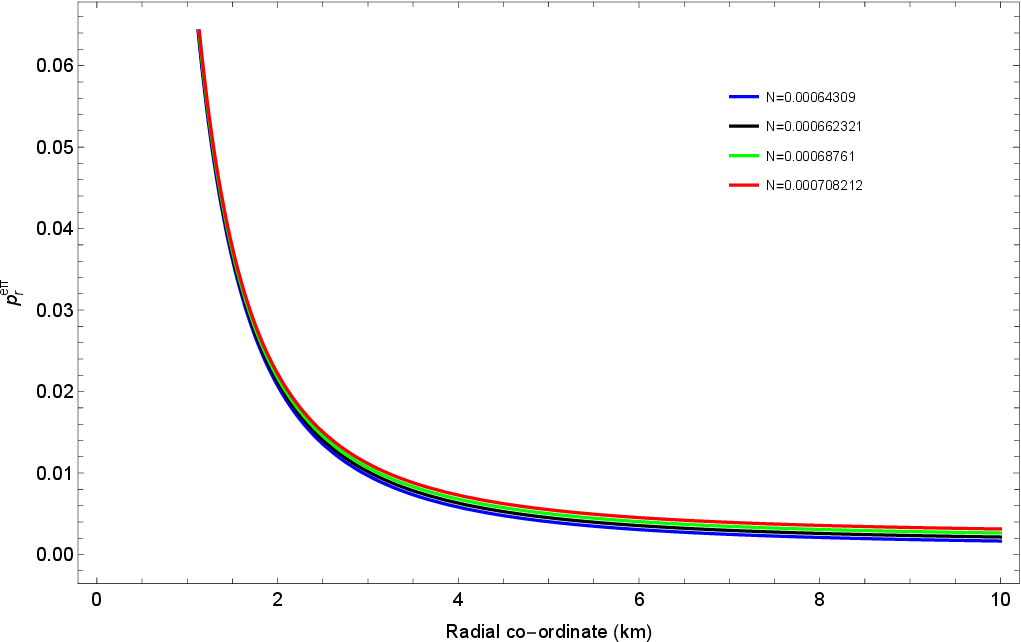}
\caption{Plot of $ p_{r}^{eff} $ versus radial co-ordinate for $ N = 0.00064309, N = 0.000662321, N= 0.00068761, N = 0.000708212 $ }\label{2}
\end{figure}\\
\begin{figure}[ht!]
\centering
\includegraphics[scale=0.5]{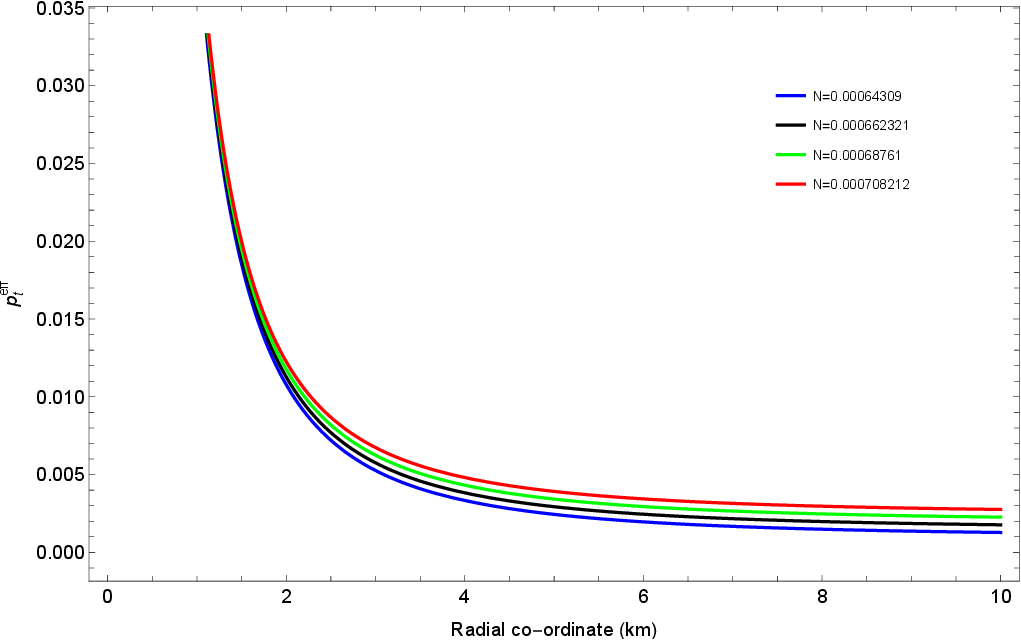}
\caption{Plot of $ p_{t}^{eff} $ versus radial co-ordinate for $ N = 0.00064309, N = 0.000662321, N= 0.00068761, N = 0.000708212 $ }\label{3}
\end{figure}\\
Figures (\ref{1}), (\ref{2}) and (\ref{3}) visually represent the results for the effective density, effective radial pressure and effective transversal pressure respectively under various parameter values. The figures clearly illustrate a smooth and decreasing trend for all the plotted quantities, which approach zero at the star's surface.\\

\section{Physical attributes of the model}\label{sec4}

\hspace{0.8cm}The effective central density $ \rho^{eff} $ at the core is provided by\\
\begin{equation}\label{34}
\rho_{c}^{eff} = \rho^{eff}(r = 0) = \frac{9N}{7 \pi}
\end{equation}
and also we have
\begin{equation}\label{35}
\frac{d\rho^{eff}}{dr} = \frac{N(4Nr + 4N^{2}r^{3})}{7\pi (1 + Nr^{2})^{3}} - \frac{6N^{2}r(9 + 2Nr^{2} + N^{2}r^{4})}{7\pi (1 + Nr^{2})^{4}}
\end{equation}
\begin{eqnarray}\label{36}
\frac{d^{2}\rho^{eff}}{dr^{2}} &=& \frac{N(4Nr + 12N^{2}r^{2})}{7\pi (1 + Nr^{2})^{3}} - \frac{12N^{2}r(4Nr + 4N^{2}r^{3})}{7\pi (1 + Nr^{2})^{4}}
\nonumber \\
&& \frac{48N^{3}r^{2}(9 + 2Nr^{2} + N^{2}r^{4})}{7\pi (1 + Nr^{2})^{5}} - \frac{6N^{2}(9 + 2Nr^{2} + N^{2}r^{4})}{7\pi (1 + Nr^{2})^{4}}
\end{eqnarray}\\
and at the core we have\\
\begin{equation}\label{37}
\frac{d^{2}\rho^{eff}}{dr^{2}}\big|_{(r=0)} = -\frac{50N^{2}}{7 \pi}.
\end{equation}\\
\begin{figure}[ht!]
\centering
\includegraphics[scale=0.5]{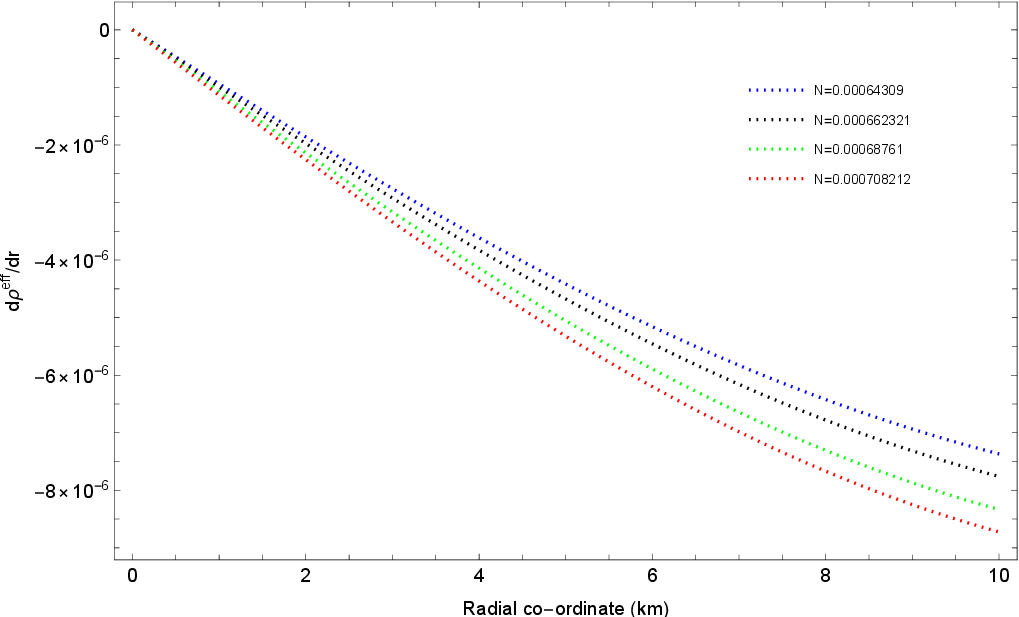}
\caption{$ \frac{d\rho^{eff}}{dr} $ against radial co-ordinate for different parameter values}\label{4}
\end{figure}\\
\begin{figure}[ht!]
\centering
\includegraphics[scale=0.5]{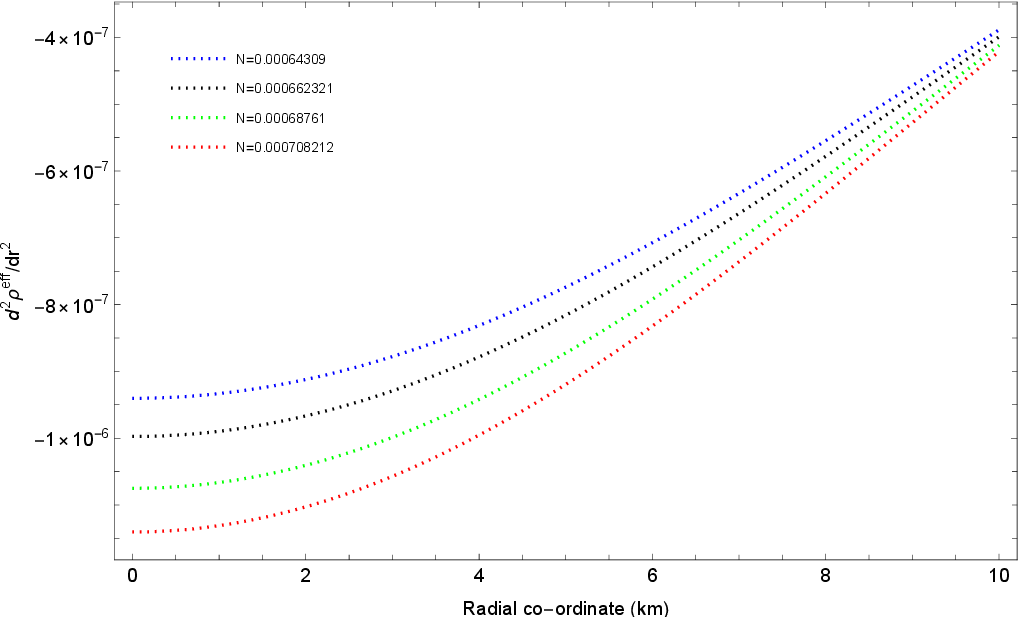}
\caption{$ \frac{d^{2}\rho^{eff}}{dr^{2}} $ against radial co-ordinate for different parameter values}\label{5}
\end{figure}\\
Thus, we confirm that the effective density is well-behaved(regular) at the star's center. Figure (\ref{4}) presents the plot of $ \frac{d\rho^{eff}}{dr} $ while figure (\ref{5}) depicts the behavior of $ \frac{d^{2}\rho^{eff}}{dr^{2}} $. Their smoothness and physically viable nature are evident from the plots. Equation (\ref{37}) confirms that the effective density has a maxima at the core $ (r=0) $. Now from the expression for $ p_{r} $ in equation (\ref{32}), we obtain\\
\begin{equation}\label{38}
\frac{dp_{r}^{eff}}{dr} = -\frac{44Nr - 20N^{2}r^{3}}{28 \pi (r + Nr^{3})^{2}} - \frac{(1 + 3Nr^{2})(7 - 22Nr^{2} - 5N^{2}r^{4})}{14 \pi (r + Nr^{3})^{2}}
\end{equation}\\
\begin{figure}[ht!]
\centering
\includegraphics[scale=0.5]{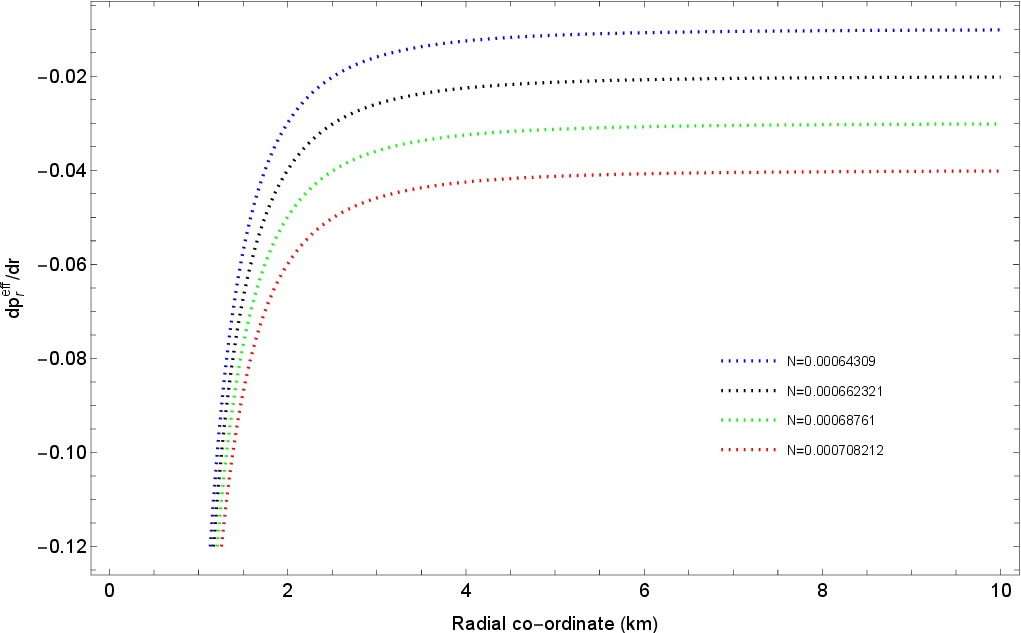}
\caption{$ \frac{dp_{r}^{eff}}{dr} $ against radial co-ordinate for $ N = 0.00064309, N = 0.000662321, N= 0.00068761, N = 0.000708212 $}\label{6}
\end{figure}\\
Figure (\ref{6}) displays the profile of $ \frac{dp_{r}^{eff}}{dr} $, which is clearly shown to be less than $ 0 $ as expected. The degree of pressure anisotropy (anisotropic factor) for our model then is defined as\\
\begin{equation}\label{39}
\triangle = p_{t}^{eff} - p_{r}^{eff} = \frac{-7-21Nr^{2} + 59N^{2}r^{4} + 9N^{3}r^{6}}{56 \pi r^{2}(1 + Nr^{2})^{3}}
\end{equation}
and the anisotropic force as\\
\begin{equation}\label{40}
\frac{2\triangle}{r} = \frac{-7-21Nr^{2} + 59N^{2}r^{4} + 9N^{3}r^{6}}{28 \pi r^{2}(r + Nr^{3})^{3}}.
\end{equation}\\
\begin{figure}[ht!]
\centering
\includegraphics[scale=0.5]{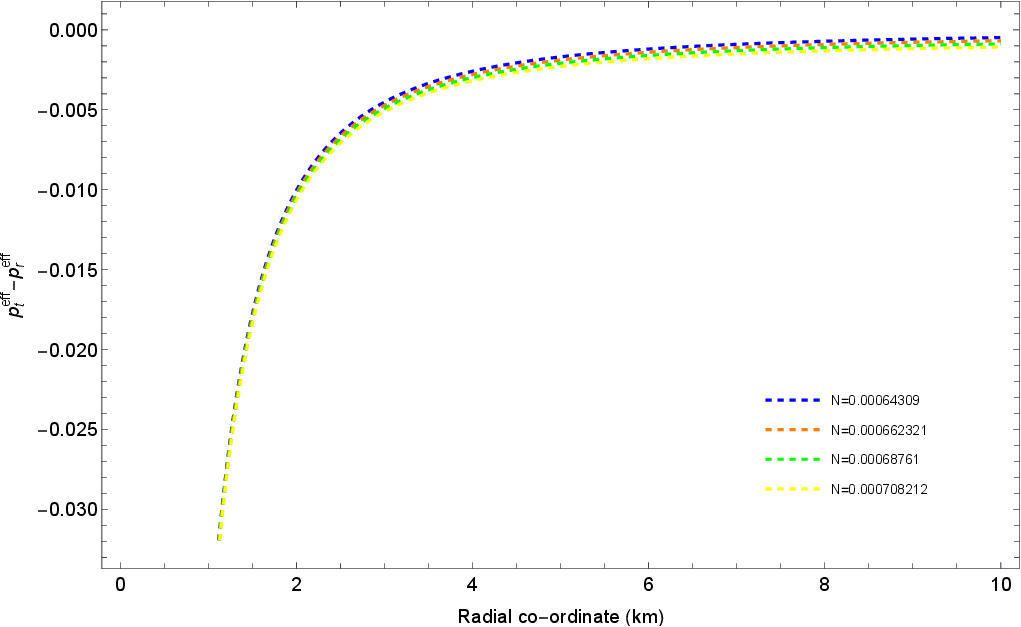}
\caption{Anisotropic factor against radial co-ordinate}\label{7}
\end{figure}\\
The negative anisotropic factor shown in figure (\ref{7}) indicates that $ p_{t}^{eff} < p_{r}^{eff} $. We can therefore conclude that the observed attractive force is due to combined effect of the star's OM and the surrounding quintessence field.\\

\section{External spacetime and the boundary constraints}\label{sec5}

\hspace{0.8cm}To explore the boundary conditions we will use the star property that it is surrounded by a vacuum spacetime and the interior is smoothly matched to the exterior Schwarzschild vacuum solution. The constants $ A $ and $ N $ are fixed by applying this boundary constraints at the hypersurface, ensuring a smooth match between the internal and the external metrics. The interior solution for the quintessence star will be joined to the exterior exactly at the stellar surface $( r= R )$, which lies beyond the event horizon $ (R > 2M )$. The geometry of the spacetime outside the star is described by the exterior metric as\\
\begin{equation}\label{41}
ds^{2} = -\big(1 - \frac{2\textsf{M}}{r}\big)dt^{2} + \frac{1}{\big(1 - \frac{2\textsf{M}}{r}\big)}dr^{2} + r^{2}(d \theta^{2} + \sin^{2}\theta d \phi^{2})
\end{equation}\\
with $ \textsf{M} $ = total gravitational mass combined within. The provided boundary conditions allow us to fix the values for the parameters $ A $ and $ N $ from below as\\
\begin{equation}\label{42}
1 - \frac{2\textsf{M}}{R} = A^{2}R^{2}
\end{equation}
\begin{equation}\label{43}
\frac{1}{1 - \frac{2\textsf{M}}{R}} = \frac{7 + 14NR^{2} + 7N^{2}R^{4}}{7 - 10NR^{2} - N^{2}R^{4}}.
\end{equation}\\

\section{Tolman - Oppenheimer - Volkoff ( TOV ) equation}\label{sec6}

\hspace{0.8cm}To analyse the stability of our compact star model against various internal forces, we use the generalised TOV equation by following the approach established in \cite{ST89}, which is expressed as\\
\begin{equation}\label{44}
-\frac{M_{G}(\rho + p_{r})}{r^{2}}e^{\frac{n-m}{2}} - p_{r}' + \frac{2(p_{t} - p_{r})}{r} = 0
\end{equation}\\
with $ M_{G} $ = effective gravitational mass. The expression obtained by starting with the modified Tolman - Whittaker  formulation as \cite{ST90}
\begin{equation}\label{45}
M_{G}(r) = \frac{r^{2}}{2}e^{\frac{n-m}{2}}n'.
\end{equation}\\
As a result, from equation (44) we have
\begin{equation}\label{44}
-\frac{(\rho + p_{r})}{2}n' - p_{r}' + \frac{2(p_{t} - p_{r})}{r} = 0.
\end{equation}\\
This equation describes how the equilibrium achieved by the simultaneous action of gravitational ($ F_{G} $), hydrostatic ($ F_{H} $) and anisotropic forces ($ F_{A} $)as\\
\begin{equation}\label{47}
F_{G} + F_{H} + F_{A} = 0
\end{equation}\\
where\\
\begin{equation}\label{48}
F_{G} = -\frac{n'}{2}(\rho^{eff} - p_{r}^{eff})
\end{equation}
\begin{equation}\label{49}
F_{H} = -\frac{dp_{r}^{eff}}{dr}
\end{equation}
\begin{equation}\label{50}
F_{A} = \frac{2(p_{t}^{eff} - p_{r}^{eff})}{r}.
\end{equation}
\begin{figure}[ht!]
\centering
\includegraphics[scale=0.5]{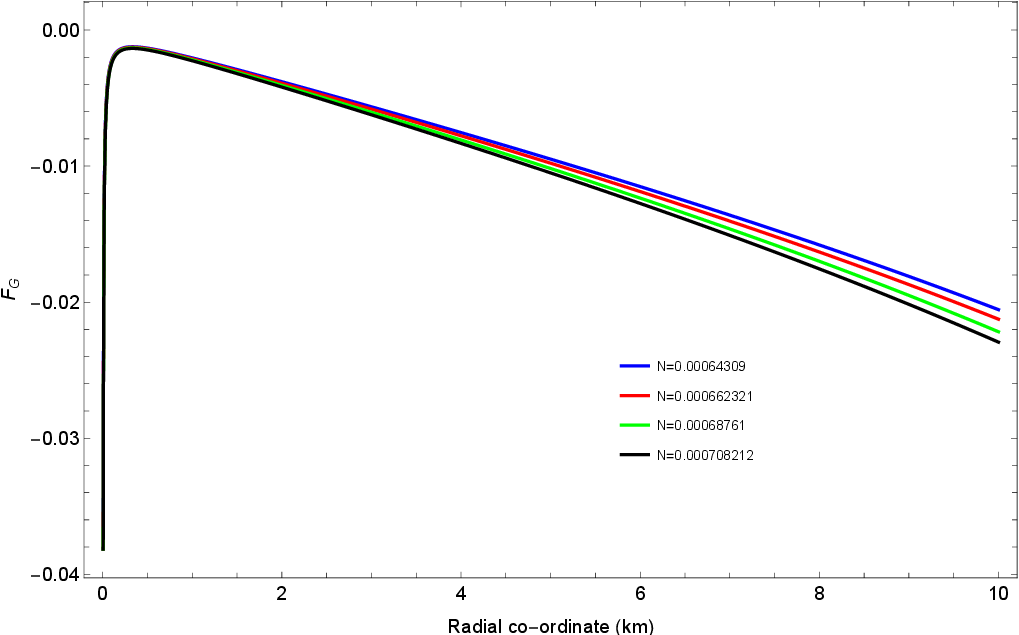}
\caption{$ F_{G} $ w.r.t radial co-ordinate}\label{8}
\end{figure}\\
\begin{figure}[ht!]
\centering
\includegraphics[scale=0.5]{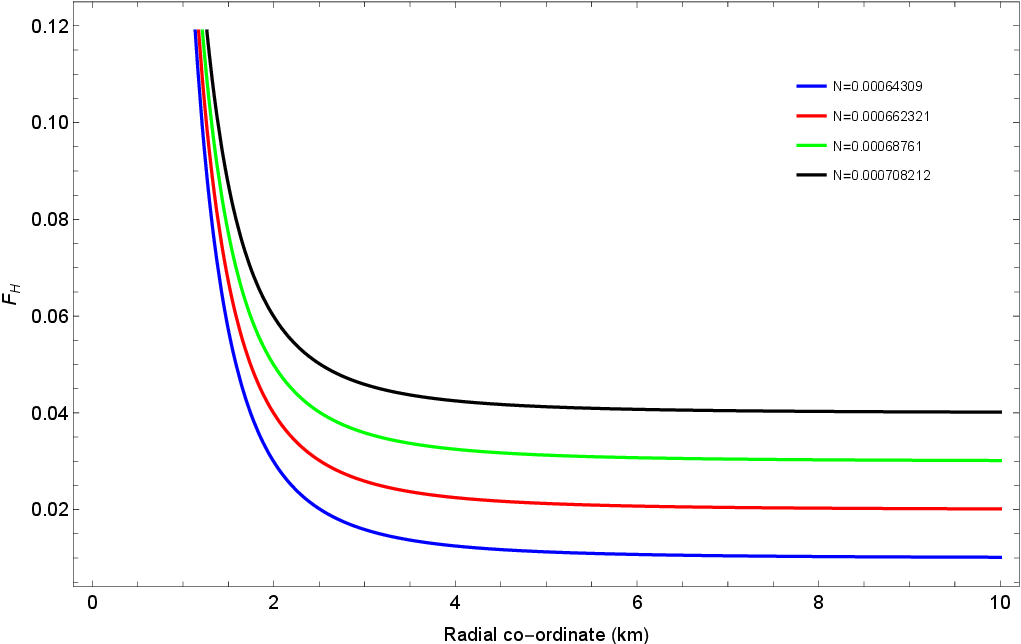}
\caption{$ F_{H} $ w.r.t radial co-ordinate}\label{9}
\end{figure}\\
\begin{figure}[ht!]
\centering
\includegraphics[scale=0.5]{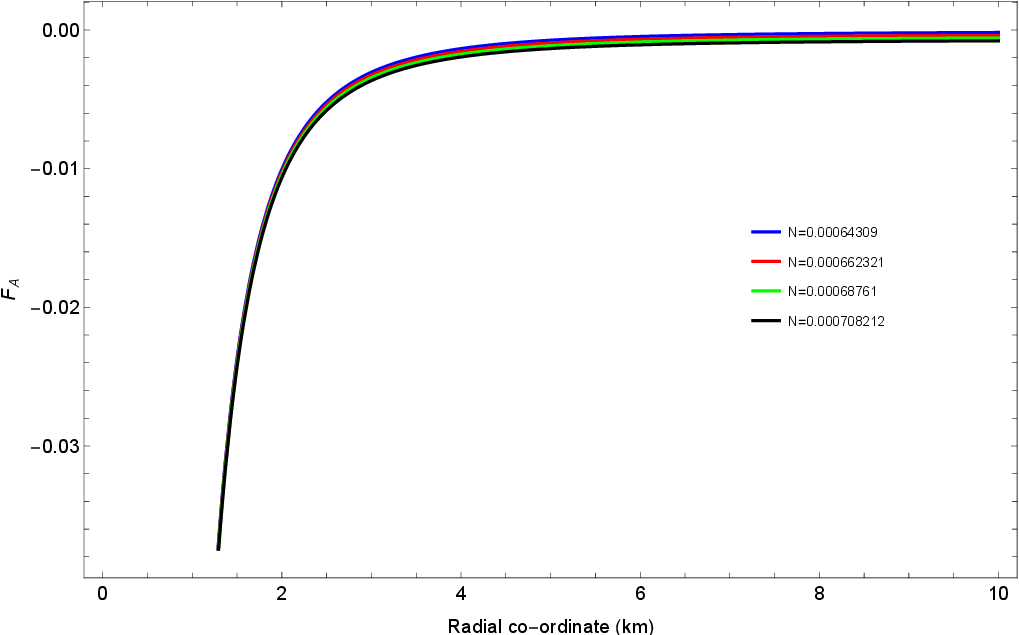}
\caption{$ F_{A} $ w.r.t radial co-ordinate}\label{10}
\end{figure}\\
The behavior of $ F_{G},  F_{H} $ and $  F_{A} $ are displayed in figures (\ref{8}), (\ref{9}) and (\ref{10}) respectively. The force profiles confirm that  $ F_{H} $ perfectly counterbalances the effects of  $ F_{G} $ and $  F_{A} $ together ensuring that our compact stellar model is in a state of stable equilibrium.\\

\section{Energy conditions}\label{sec7}

\hspace{0.8cm}The energy momentum describes the distribution of energy momentum, stress contributed by matter of non-gravitational forces in a given region of spacetime. The EFEs allow physicists to easily include the effects of diverse matter in their models. The energy conditions define the standard constraints, confirming that the properties assigned to matter and non-gravitational fields across different physical scenarios. These energy conditions are powerful constraints which discard solution to the EFEs that describe unrealistic or non-physical solutions. To fully understand the physical behavior of a star made of anisotropic quintessence, we absolutely must verify that its theoretical model adheres to the energy conditions. To ensure that the energy conditions viz.,
\begin{itemize}
 
\item \textbf{Null energy condition (NEC)}
\item \textbf{Weak energy condition (WEC)}
\item \textbf{Strong energy condition (SEC)}
\item \textbf{Dominant energy condition (DEC)}
\
\end{itemize}
are met all the specified inequalities below must be true at every single point within the spherical fluid object. They are given as\\
\begin{equation}\label{51}
NEC : \rho^{eff} + p_{r}^{eff} \geq 0, \hspace{0.5cm} \rho^{eff} + p_{t}^{eff} \geq 0
\end{equation}
\begin{equation}\label{52}
WEC : \rho \geq 0 \hspace{0.5cm} \rho^{eff} + p_{r}^{eff} \geq 0, \hspace{0.5cm} \rho^{eff} + p_{t}^{eff} \geq 0
\end{equation}
\begin{equation}\label{53}
SEC : \rho^{eff} + p_{r}^{eff} +2 p_{t}^{eff} \geq 0
\end{equation}
\begin{equation}\label{54}
DEC : \rho^{eff} - \mid p_{r}^{eff} \mid \geq 0, \hspace{0.5cm} \rho^{eff} - \mid p_{t}^{eff} \mid \geq 0.
\end{equation}\\
\begin{figure}[ht!]
\centering
\includegraphics[scale=0.5]{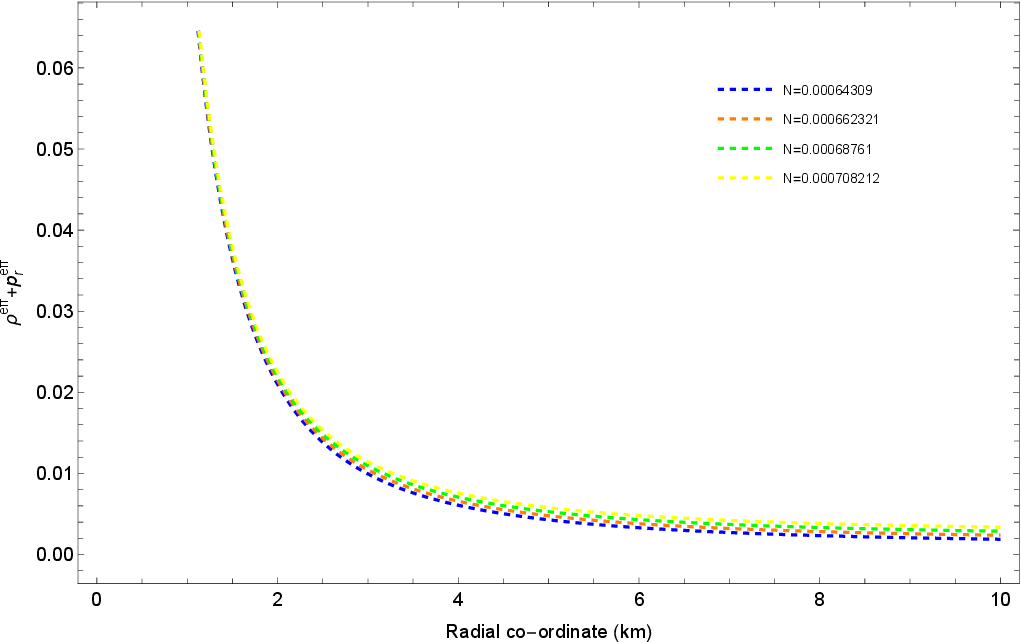}
\caption{$ NEC $ versus radial co-ordinate}\label{11}
\end{figure}\\
\begin{figure}[ht!]
\centering
\includegraphics[scale=0.5]{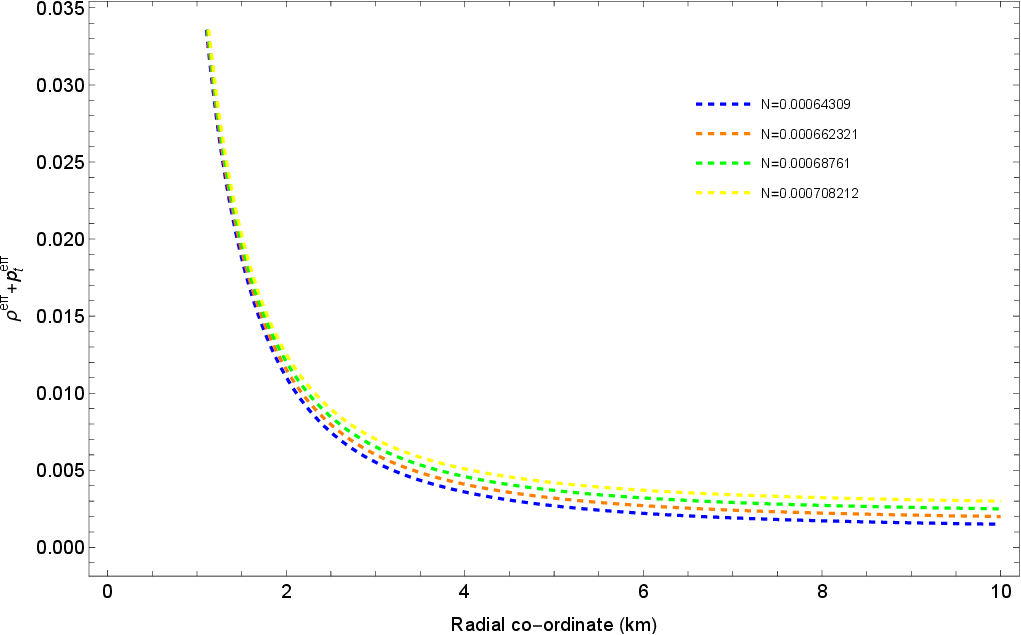}
\caption{$ NEC $ versus radial co-ordinate}\label{12}
\end{figure}\\
\begin{figure}[ht!]
\centering
\includegraphics[scale=0.5]{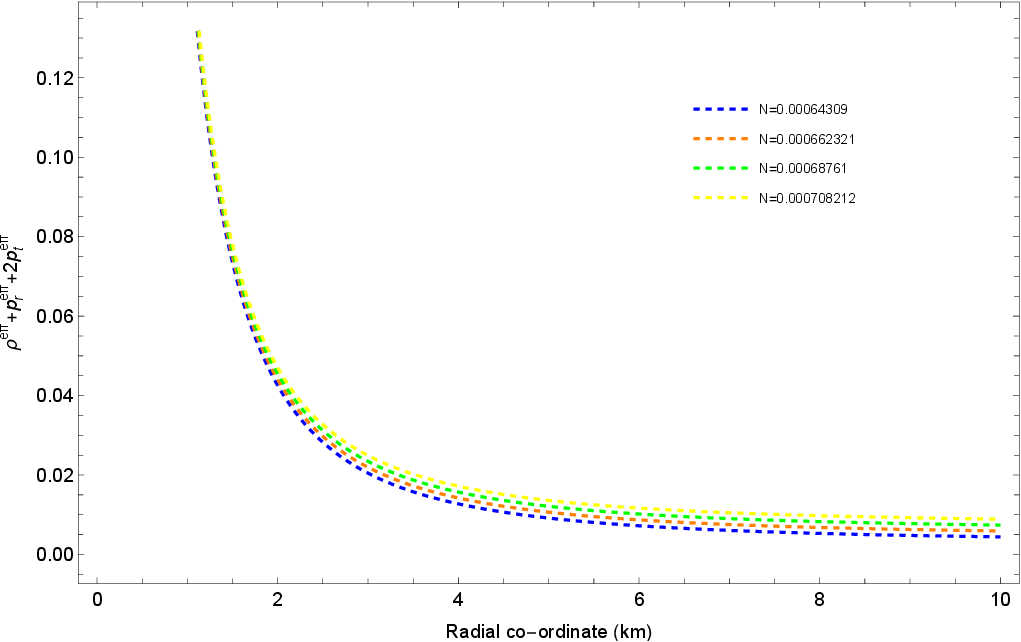}
\caption{$ SEC $ versus radial co-ordinate}\label{13}
\end{figure}\\
\begin{figure}[ht!]
\centering
\includegraphics[scale=0.5]{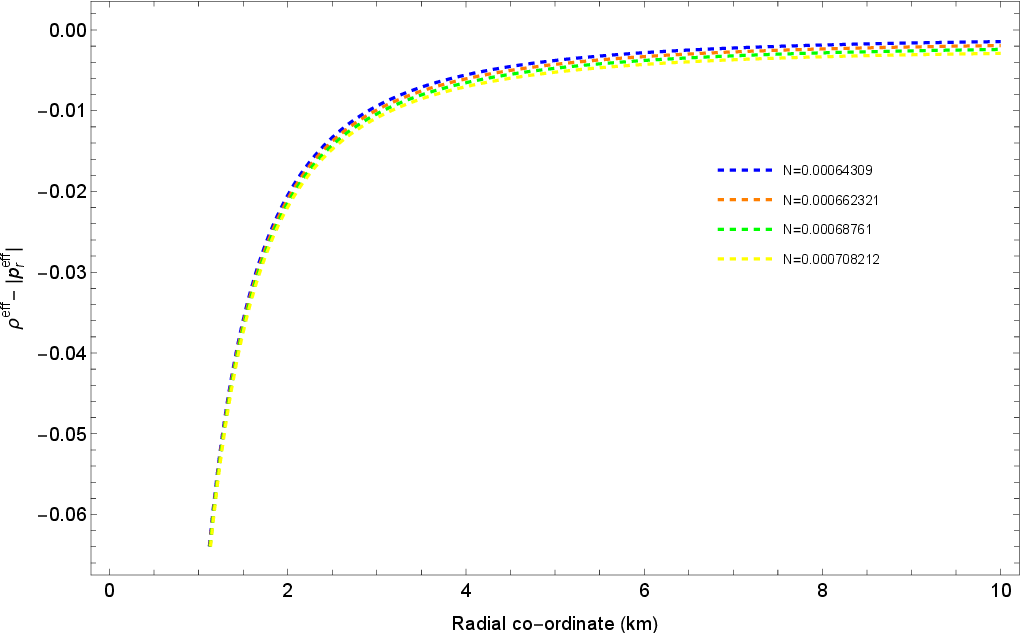}
\caption{$ DEC $ versus radial co-ordinate}\label{14}
\end{figure}\\
\begin{figure}[ht!]
\centering
\includegraphics[scale=0.5]{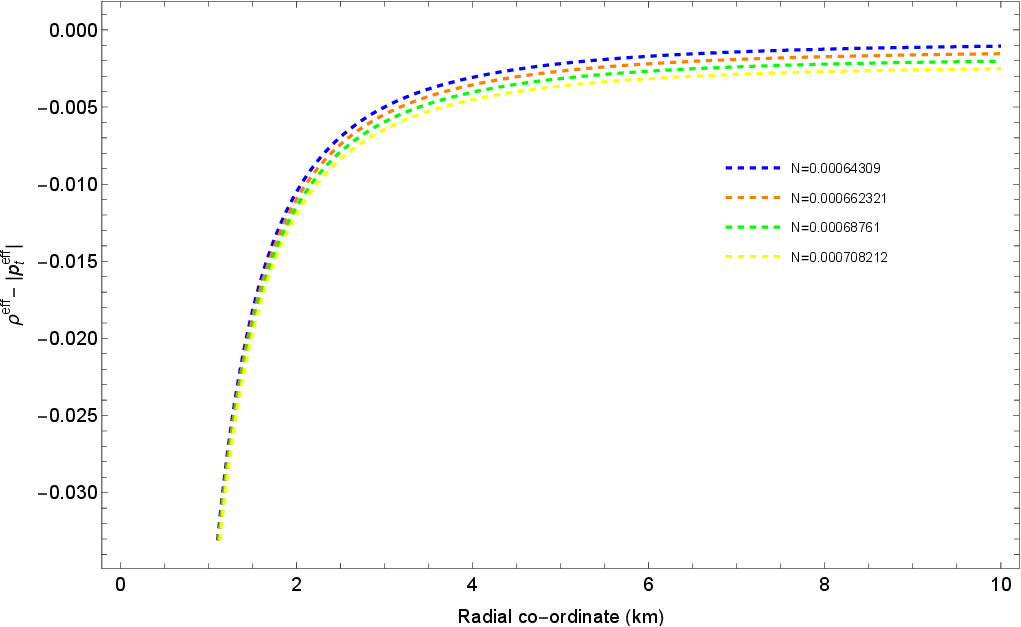}
\caption{$ DEC $ versus radial co-ordinate}\label{15}
\end{figure}\\
Our model meets the $ NEC, WEC $ and $ SEC $ as demonstrated in the figures (\ref{11}-\ref{15}). The satisfaction of $ SEC $ suggests that our spacetime is stable against gravitational collapse, leasing to the conclusion that our spacetime model is free of black holes. Being a DE star it violates the $ DEC $, as its internal negative pressure or scalar-field dynamics allow energy densities and fluxes that do not behave like normal matter - they can imply superluminal energy transport or exotic spacetime curvature effects.\\

\section{Some stellar features}\label{sec8}

\hspace{0.8cm}This section will focus on the key properties of our quintessence star model, particularly within the framework of DP metric and using the conformal symmetry.\\

\subsection{Mass radius relation}

\hspace{0.5cm}We have the effective mass for the static, anisotropic spherical fluid distribution, from the density profile outlined in equation (\ref{31}) as\\
\begin{equation}\label{55}
\breve{M} = \int_{0}^{R} 4 \pi r^{2} \rho^{eff} dr = \frac{4 NR^{3}(3 +  NR^{2})}{7(1 + NR^{2})^{2}}. 
\end{equation}\\
\begin{figure}[ht!]
\centering
\includegraphics[scale=0.5]{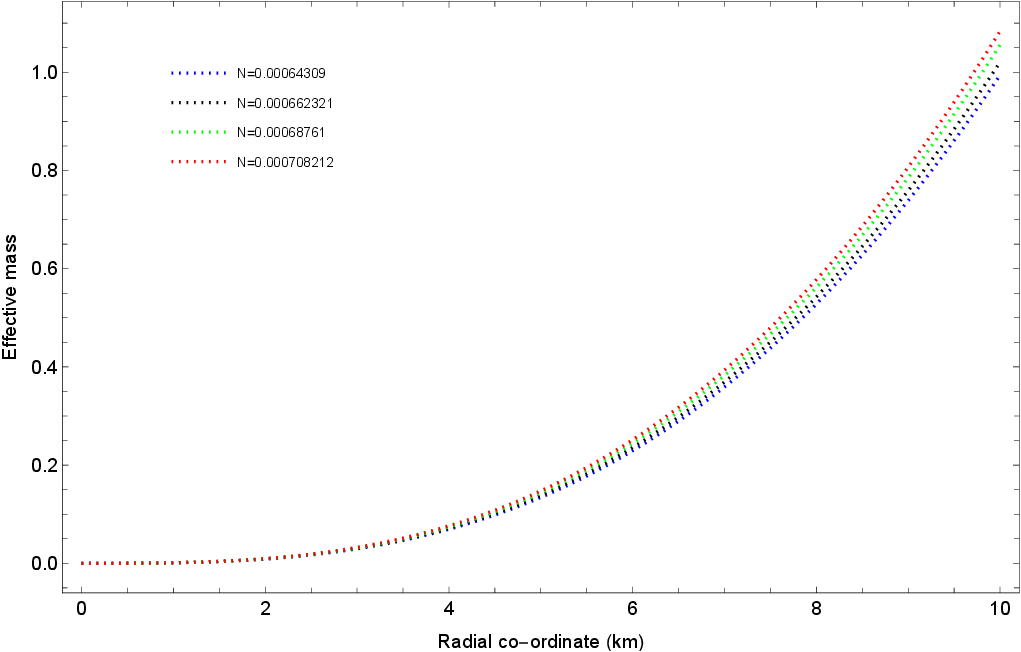}
\caption{Graph of effective mass against radial co-ordinate}\label{16}
\end{figure}\\
The graphical display is presented in the plot of figure (\ref{16}). The figure clearly shows that the mass function is regular at the center (tends to $ 0 $ as $ r $ tends to $ 0 $), it is also a positive and monotonically increasing function which aligns with the expected physical properties.\\

\subsection{Compactness}

\hspace{0.5cm}The $ \frac{\breve{M}}{r} $ relation, known as the compactness factor provides a mean to categorize various types of stellar objects. For our model, we have the compactness factor $ \hat{u}(r) $ \cite{ST} as\\
\begin{equation}\label{56}
\hat{u}(r) = \frac{\breve{M}}{r} = \frac{4 Nr^{2}(3 +  Nr^{2})}{7(1 + Nr^{2})^{2}}.
\end{equation}
\begin{figure}[ht!]
\centering
\includegraphics[scale=0.5]{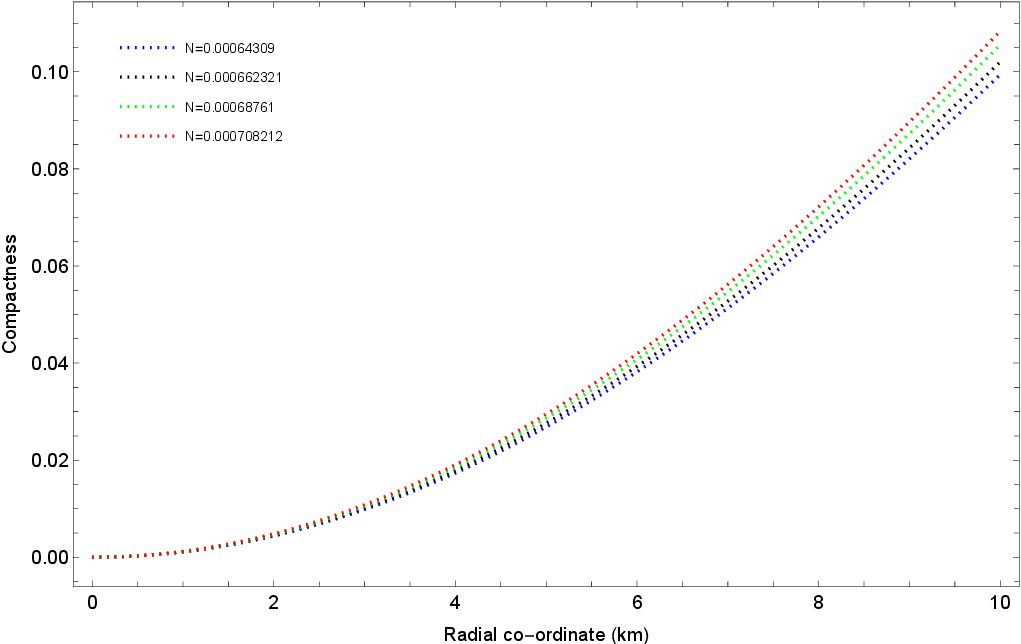}
\caption{Graph of the compactness factor against radial co-ordinate}\label{17}
\end{figure}\\
The graph in figure (\ref{17}) shows that the compactness factor steadily rises as the radius increases. Since the maximum value remains within the Buchdahl limit of $ 4/9 $, this graphical behavior suggests that our model describes a characteristics configuration of an ultra-dense stellar object.\\

\subsection{Surface redshift and gravitational redshift}

\hspace{0.5cm}For our model, the surface redshift $ \texttt{z}_{s} $ and gravitational redshift $ \texttt{z} $ are given by\\
\begin{equation}\label{57}
\texttt{z}_{s} = \frac{1}{(1 - 2\hat{u}(r))^{\frac{1}{2}}} - 1
\end{equation}
\begin{equation}\label{58}
\texttt{z} = e^{-\frac{m(r)}{2}} - 1.
\end{equation}
\begin{figure}[ht!]
\centering
\includegraphics[scale=0.5]{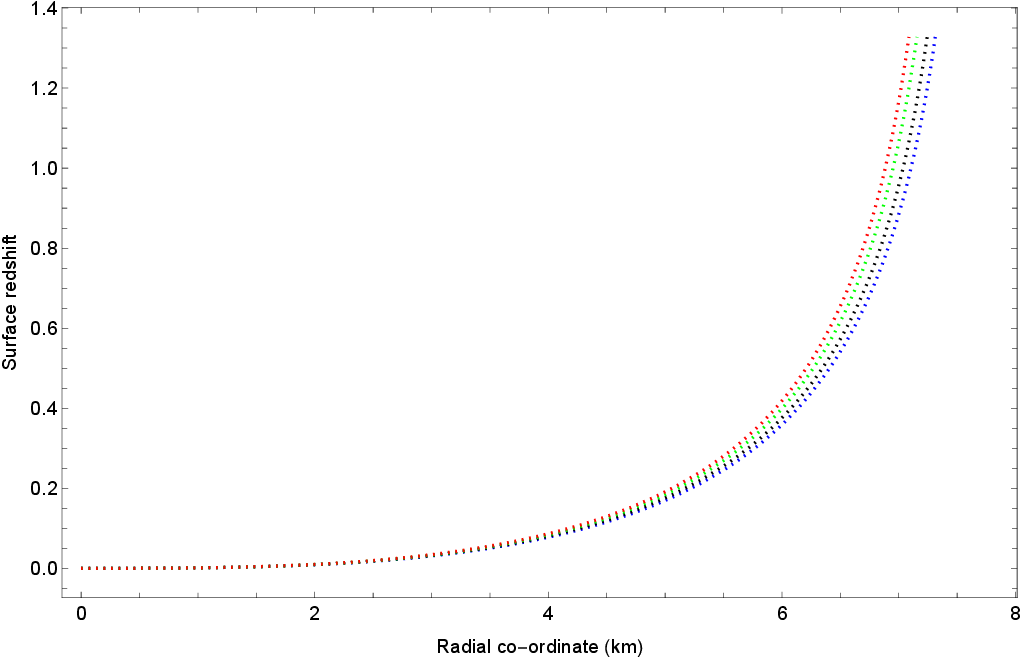}
\caption{Variation of the redshift function w.r.t the radial co-ordinate}\label{18}
\end{figure}\\
The surface gravity, controlled by the stellar object's mass and radius, is the key determinant of the surface redshift. The gravitational redshift describes the effect where photons (electromagnetic waves) are observed to decrease in energy when they move away from a source of gravity. This energy loss corresponds to an increase in their observed wavelength. Under the condition of a vanishing cosmological constant, $ \texttt{z}_{s} $ is constrained to be less than or equal to $ 2 $ for an isotropic stellar model \cite{ST17}. Subsequent research established that for a star with anisotropy, the limit substantially reaches to $ 5 $ \cite{ST18}. The initial theoretical constraint was subsequently adjusted, establishing 5.211 as the new calculated maximum value for the redshift function \cite{ST19}. As figure (\ref{18}) illustrates that the model's behavior consistently satisfy the specified range in our case. The surface redshift is maximal at the surface and decreases towards the center. This occurs because the small increase in radius relative to mass increase leads to a greater surface gravity. We have $ \texttt{z}(R) = \texttt{z}_{s}(R) $, when measured precisely at the surface of the stellar object. The graphical representation also confirms that the redshift function exhibits regular behavior, being positive, finite and singularity free across the model's structure. All of these characteristics offer a strong support to the validity of our model.\\

\section{Stability analysis}\label{sec9}

\hspace{0.8cm}Stability is a critical prerequisite for determining if a stellar structure model represents a physically achievable star. We will take the help of the radial and transverse sound speed to assess the stability, where the mathematical forms are provided via\\
\begin{equation}\label{59}
V_{r}^{2} = \frac{dp_{r}}{d \rho} = \varpi < 1
\end{equation}
\begin{eqnarray}\label{60}
V_{t}^{2} & = & \frac{dp_{t}}{d \rho} = \frac{1}{2(-7 - 28Nr^{2} - 82N^{2}r^{4} + 36N^{3}r^{6} + N^{4}r^{8})}
\nonumber \\
&& \Big(2(-7 - 28Nr^{2} + 88N^{2}r^{4} + 16N^{3}r^{6} + 3N^{4}r^{8}) + \varpi(-7 - 28Nr^{2} + 258N^{2}r^{4} - 4N^{3}r^{6} + 5N^{4}r^{8}) + 
\nonumber \\
&& 3w_{Q}(-7 - 28Nr^{2} + 2N^{2}(9 + 50\varpi)r^{4} + 4N^{3}(1 + 2\varpi)r^{6} + N^{4}(5 + 4\varpi)r^{8})\Big).
\end{eqnarray}\\
\begin{figure}[ht!]
\centering
\includegraphics[scale=0.5]{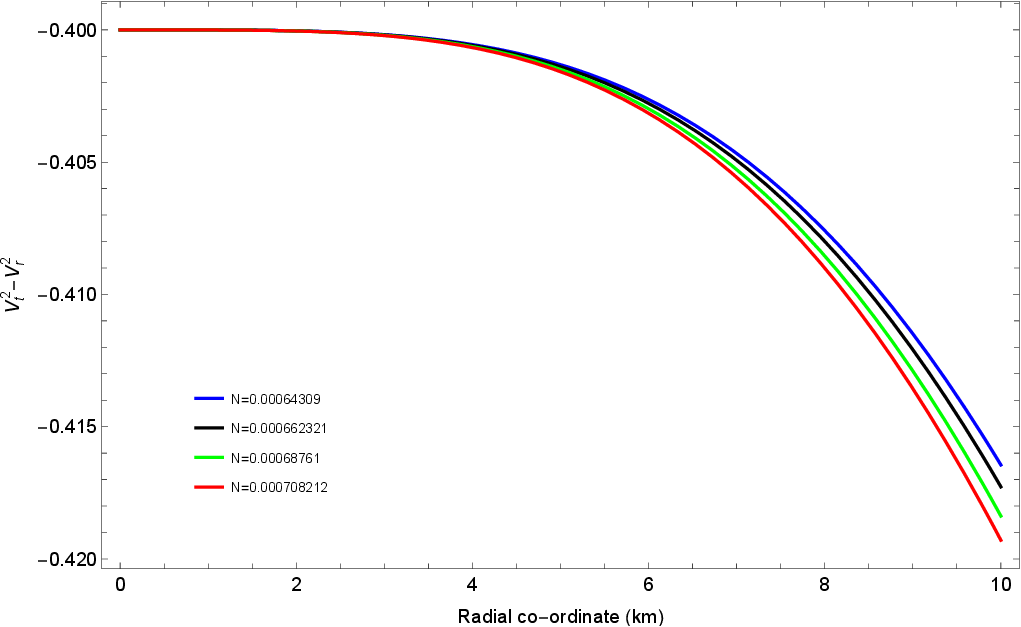}
\caption{Plot of $ V_{t}^{2} -  V_{r}^{2} $ vs radial co-ordinate}\label{19}
\end{figure}\\
\begin{figure}[ht!]
\centering
\includegraphics[scale=0.5]{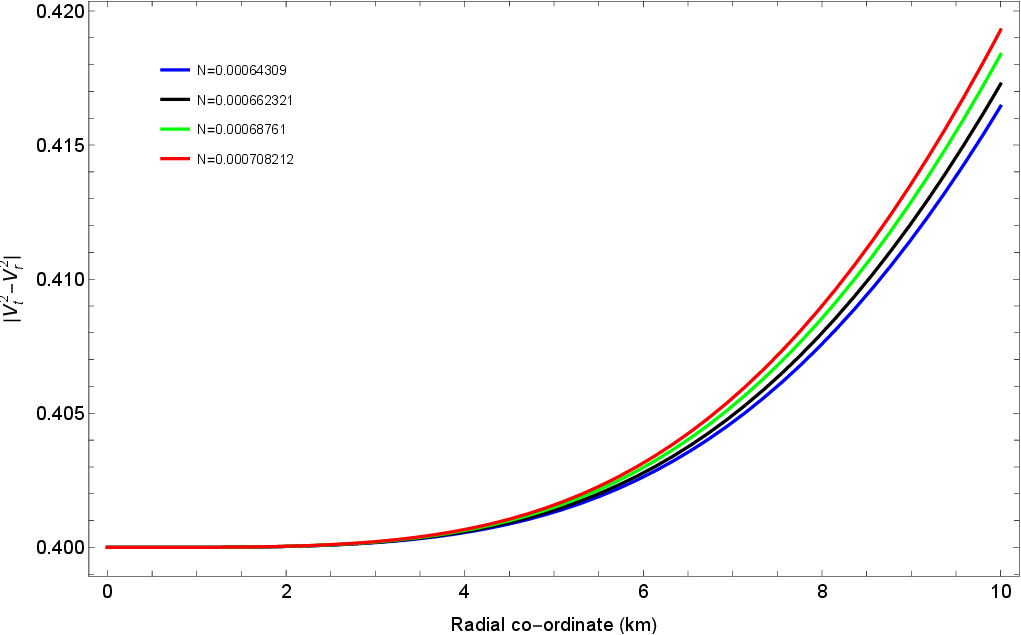}
\caption{Plot of $ | V_{t}^{2} -  V_{r}^{2} | $ vs radial co-ordinate}\label{20}
\end{figure}\\
A new technique was developed to study how matter is distributed and it includes the introduction of the cracking concept \cite{RG53,RG54,RG55}. Potential stability is assessed by evaluating the difference in sound speeds (radial vs transversal) across various stellar regions. Stability is often achieved in stellar regions where the transverse propagation speed of sound is slower than the radial speed. The practical success of the model rests on demonstrating that the resulting structure is stable. The visual demonstration in figure (\ref{19}) confirms that the model meets the condition where $ V_{t}^{2} -  V_{r}^{2} < 0 $. Since the graph in figure (\ref{20}) indicates that $ 0 \leq | V_{t}^{2} -  V_{r}^{2} | \leq 1 $, the crucial stability condition is met, validating the proposed star model.\\

\section{Discussions and conclusion}\label{sec10}

\hspace{0.8cm}This research paper introduces a novel model for a compact star. This model utilizes the D-P metric and is constructed under the assumption that it admits a conformal killing vector. Crucially, the model incorporates a quintessence field which is characterized by a specific parameter $ w_{Q} (-1 < w_{Q} < -\frac{1}{3}) $. We have achieved a physically acceptable solution in this regard. Our objective was to conduct a more detailed investigation of the physical properties of the model after successfully generating an exact solution. The summary of our findings are given below:\\
\begin{itemize}
\item \textbf{Density and pressure} - The model ensures that the effective density remains regular at the star's core. The monotonic decreasing nature that is observed in both the density formation and the radial pressure profiles confirm that these physical quantities are maximal at the stellar center and is minimal at the surface.
\item \textbf{TOV equations} - The model's stability is confirmed by its adherence to the requirements of the generalized TOV equations. As illustrated in figures (\ref{8}), (\ref{9}) and (\ref{10}), the combined influence of the forces $ F_{G} $ and $ F_{A} $ is effectively counteracted by the effect of the force $ F_{H} $.
\item \textbf{Energy conditions} - The model's satisfaction of the $ NEC, WEC $ and $ SEC $ and the violation of the $ DEC $ serve as a compelling reason for its validity, as this combination of results precisely characterizes a DE star and it is clearly evident from the plots of figures (\ref{11}-\ref{15}).
\item \textbf{Mass radius relation} - The variation of the effective mass function w.r.t the radial co-ordinate is presented in figure (\ref{16}). The variation confirms that the mass function is well-behaved at the center (as $ r \rightarrow 0, \breve{M} \rightarrow 0 $). In the present work, we can also see that the mass-radius ratio clearly lies within the accepted Buchdahl limit.
\item \textbf{Compactness and redshift} - Figures (\ref{17}) and (\ref{18}) visually present the changes in compactness factor and redshift function for the model. The figures clearly demonstrate that both the functions exhibit continuity as the radial co-ordinate approaches $ 0 $, and their values increase for larger $ r $, holding true across various parameter values. Our analysis provides powerful evidence that our quintessence star candidates are physically consistent.
\item \textbf{Casuality condition} - The stability of our model is also confirmed by the fact that the radial and transverse speeds of sound are less than unity. Herrera's condition for potential model stability requires that the radial sound speed to be greater than the transverse sound speed. Figure (\ref{19}) displays the fulfillment of the condition $ V_{r}^{2} -  V_{t}^{2} > 0 $ while figure (\ref{20}) further establishes the model's physical acceptability by showing that $ | V_{t}^{2} -  V_{r}^{2} | \leq 1 $.
\end{itemize}
In conclusion, our proposed model successfully represents a physically viable, stable and singularity-free compact star model. This model incorporates a quintessence field and admits conformal motion, making it a robust representation of a highly compact star. This model perfectly qualifies as the representative scenario for analyzing the physical characteristics of quintessence star and is highly relevant for subsequent investigations.\\

\end{document}